\documentclass[
    ,final            % use final for the camera ready runs
  ]
  {aipproc}
\layoutstyle{8x11single}

\usepackage[]{natbib}

\begin{document}

\title{Studies of Hadronization Mechanisms using Pion Electroproduction in Deep Inelastic Scattering from Nuclei}

\classification{\texttt{13.60.Le, 13.87.Fh, 14.65.Bt, 25.30.Rw}}

\author{Will Brooks, Hayk Hakobyan, Cristian Pe\~na, Miguel Arratia, Constanza Vald\'es\\for the CLAS Collaboration}{
  address={Institute for Advanced Studies in Science and Engineering,
  Valpara\'iso Center for Science and Technology,
  Universidad T\'ecnica Federico Santa Mar\'ia,
  Avda. Espa\~na 1680, Casilla 110-V Valpara\'iso, Chile}
}

\begin{abstract}
Atomic nuclei can be used as spatial analyzers of the hadronization process in semi-inclusive deep inelastic scattering. The study of this process using fully-identified final state hadrons began with the HERMES program in the late 1990s, and is now continuing at Jefferson Lab. In the measurement described here, electrons and positive pions were measured from a 5 GeV electron beam incident on targets of liquid deuterium, C, Fe, and Pb using CLAS in Hall B. The broadening of the transverse momentum of positive pions has been studied in detail as a function of multiple kinematic variables, and interpreted in terms of the transport of the struck quark through the nuclear systems. New insights are being obtained into the hadronization process from these studies; and experiments of this type can be relevant for the interpretation of jet quenching and proton-nucleus collisions at RHIC and LHC. These measurements will be extended in the next few years with the approved JLab experiment E12-06-117, and later at a future Electron-Ion Collider.
\end{abstract}

\maketitle

\vspace{-0.3cm}
\section{Introduction}
\vspace{-0.2cm}
Semi-inclusive hadron production from nuclear targets can be used to study fundamental aspects of the hadronization process in QCD. Using the well-known properties of nuclei, the interactions of the propagating quarks and forming hadrons with the nuclear medium reveal the details of this process at femtometer time and distance scales. Pioneering studies of this kind using identified hadrons were first performed by the HERMES experiment beginning in the 1990's. New experiments at Jefferson Lab\cite{jlab:2002} are now beginning to provide more detailed insight into the low-energy region with orders of magnitude more luminosity and extensions to the heaviest nuclear targets.

Exploration of the low energy region takes advantage of a unique kinematic window wherein the predominant time scales of fundamental QCD processes are comparable to the dimensions of atomic nuclei. From older studies without hadron identification, as well as from the HERMES investigations with identified hadrons, effects on hadron production due to the nuclear medium diminish at higher parton energies\cite{hermes:2007}, except possibly at large $x_F$ \cite{kopel:2005}. While explorations at lower energies may yield the most information on parton propagation, they also require careful study of the basic features of the interactions, such as the roles played by current and target fragmentation, and of the potential influence of resonances. These two topics are explored below, followed by a qualitatively new observation of a $\rm \phi_{pq}$ dependence, preliminary fit results for a characteristic time in QCD, and an exploration of a direct measurement of quark energy loss. 
 
\vspace{-0.3cm} 
\section {$\rm p_T$ broadening: current fragmentation, quantum effects} 
\vspace{-0.2cm}

\vspace{-0.3cm}
\subsection{Dependence on Feynman x and on W}
\vspace{-0.2cm}
As noted above, in these studies one must characterize the nature and mechanisms of the interactions before strong conclusions about QCD processes can be drawn. Basic questions include: how can behavior due to partonic and hadronic degrees of freedom be distinguished? How is hadronization modified in the nuclear medium? Using $\rm p_T$ broadening ($\rm \Delta p_T^2$) of hadrons produced in DIS from nuclear targets, one can infer information on mean interaction strengths (e.g. the transport parameter $\rm \hat{q}$), characteristic times (e.g. the lifetime of the propagating quasi-free quark, the "production time" $\rm \tau_p$), and hadron formation mechanisms, which are the dynamical enforcement of QCD confinement.

Feynman x ($\rm x_F$), which is the normalized longitudinal momentum of the hadron in the $\gamma^*-N$ center of momentum frame, has traditionally been interpreted as distinguishing current fragmentation from target fragmentation. At high z ($\rm z_h=E_{hadron}/\nu$) $\rm x_F$  and z are correlated, but at lower $\rm z_h$ a wider range of values of $\rm x_F$ is possible, and the intercomparison of the dependence of ($\rm \Delta p_T^2$) on these two variables can provide insight into the microscopic behavior of the interaction. In Fig. 1A is shown a plot of $\rm \Delta p_T^2$ vs. $z_h$ with and without the cut $\rm x_F>0$ for Pb in one particular bin in $\rm \nu$ and $Q^2$, integrated over all other variables. What is observed is that for $z_h>0.5$, the $\rm \Delta p_T^2$ is identical within the statistical uncertainties, but for $z_h<0.5$ the broadening is increasingly due to the region $\rm x_F<0$. Interpretation of this observation depends on one's assumptions. For example, if $\rm x_F>0$ is taken as a definition of current fragmentation, it means that part of the broadening below $z_h>=0.5$ is due to the target fragmentation region: about half for $0.4<z_h<0.5$, and essentially all for $z_h<0.3$. This is the first time such a study has ever been possible, and through a systematic comparison of the multidimensional behavior to model calculations, it should be possible to gain insight into the details of hadronization dynamics for intermediate $\rm z_h$.

Initial studies of the variation of $\rm \Delta p_T^2$ on the value of W (mass of recoiling hadronic system) for W>2 GeV thus far only indicate a constant dependence within the uncertainties for $2<W<3$. If resonances in the recoiling mass system were to affect the results, it would be expected to be visible as a W dependence. These results could in principle be extended to even lower W. However, for the purposes of studying hadronization, it seems that the traditional lower limit of W>2 GeV is adequate to exclude resonance effects, as expected.

\vspace{-0.3cm}
\subsection{Dependence on $\rm \phi_{pq}$}
\vspace{-0.2cm}
The variable $\rm \phi_{pq}$ is the azimuthal angle between the lepton scattering plane and the plane defined by the direction of the virtual photon and the hadron. This variable has an important role in both polarized and unpolarized electron scattering. On very general grounds the cross section for electron-hadron scattering, including electron-nucleus scattering, depends on this variable with the form $f_\phi = constant+cos(\phi_{pq})+cos(2\phi_{pq})$, independent of process. The dependence of $\rm \Delta p_T^2$ on $\rm \phi_{pq}$ has never previously been investigated experimentally. It has been generally assumed that the multiple scattering, whether partonic or hadronic, in passing through a nucleus is correctly viewed in a classical picture that ignores any coherent effects, and that this classical picture is increasingly valid for larger and larger nuclear systems. Thus the expectation is that any phi dependence would be "smeared out" for larger nuclei. However, as can be seen in Fig. 1B, we have discovered a modulation of $\rm \Delta p_T^2$ that does not shrink with increasing nucleon number A. The curves shown are fitted with the function $f_\phi$ only for reference, to illustrate the behavior most clearly; $f_\phi$ is not the expected function in this case, in fact, classically the expected function is $g_\phi(\phi_{pq}) \approx constant$.

The detailed interpretation of this observation will require theoretical work. However, at a minimum, it implies the need for a quantum mechanical analysis. Whether this observation is somehow simply an indirect reflection of the known behavior of the cross section according to $f_\phi$, or rather whether it is a rich new source of physics insight, remains to be seen. Future experimental explorations of this phenomenon could include study of the dependence on path length through the medium, as measured through, e.g., the charged particle multiplicity, and of measurements with polarized beam.

\begin{figure}[h]
  \includegraphics[height=.3\textheight]{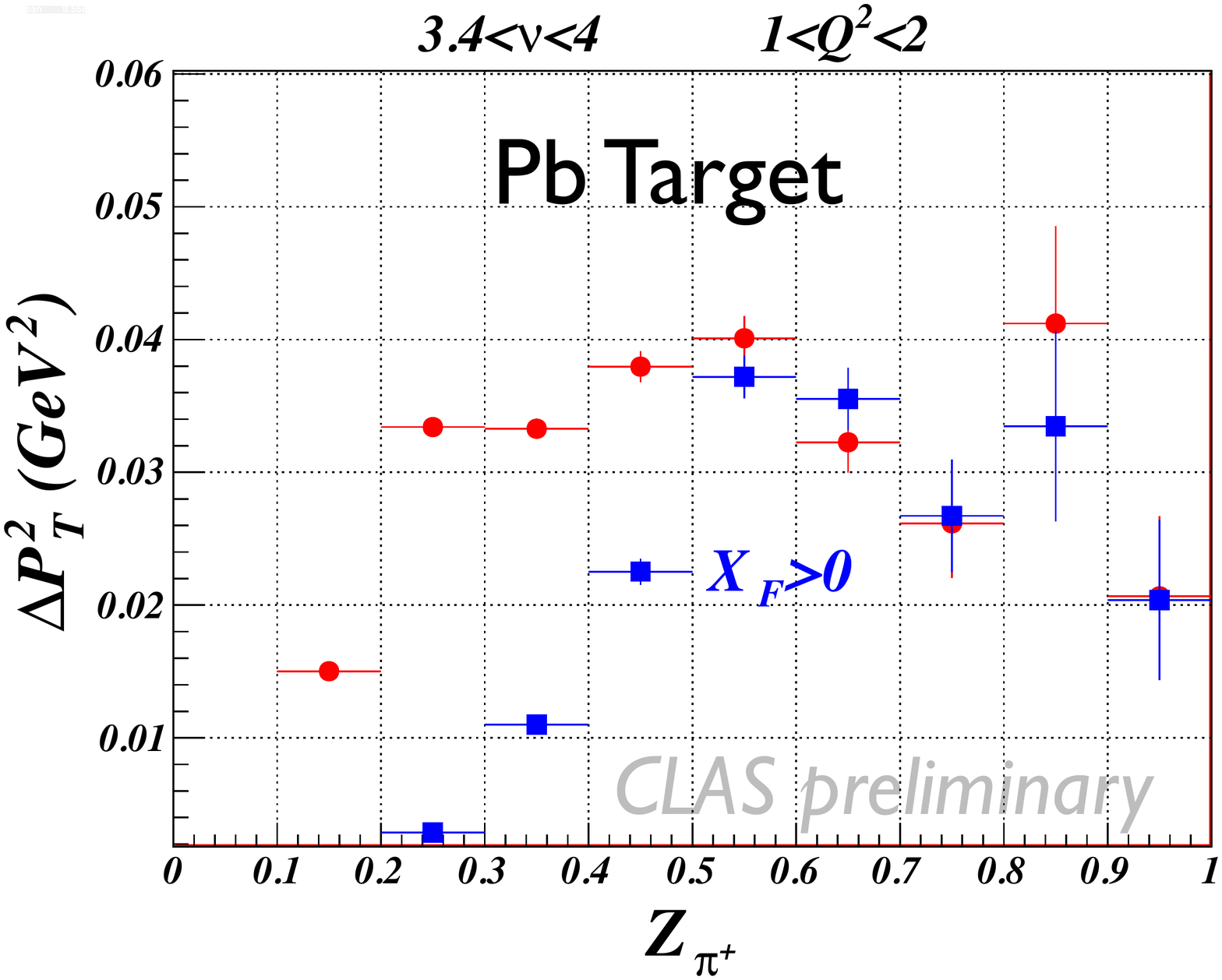}
  \includegraphics[height=.3\textheight]{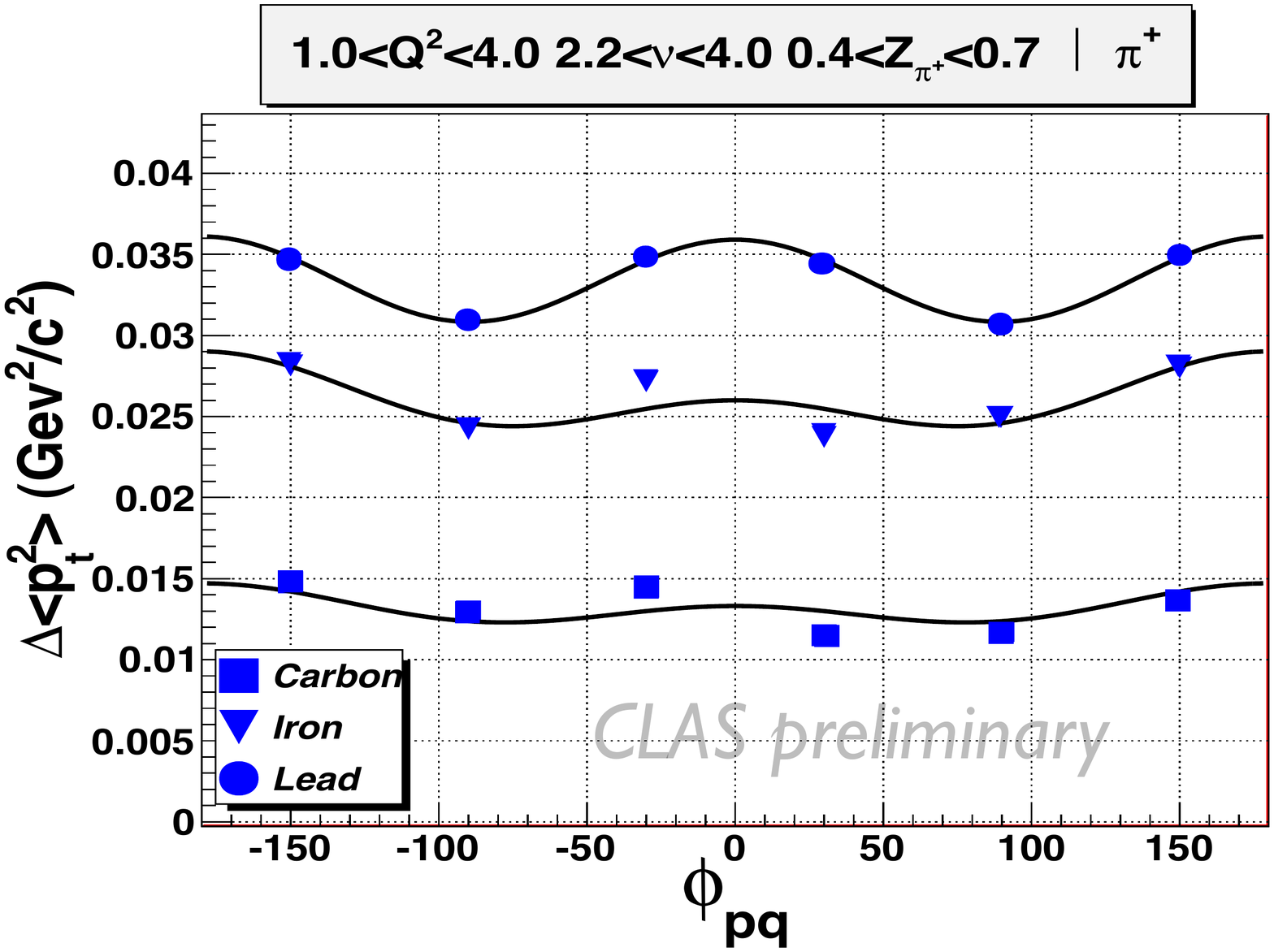}
  \caption{(Left, panel A) $\rm \Delta p_T^2$ vs. $z_h$ for $\rm \pi^+$ in Pb relative to deuterium in the bin of $\rm \nu$ and $\rm Q^2$ shown. Solid symbols are for $\rm x_F>0$ (nominally current fragmentation), while hollow symbols are integrated over all $\rm x_F$. (Right, panel B) $\rm \Delta p_T^2$ vs. $\phi_{pq}$ for $\rm \pi^+$ in C, Fe, and Pb integrated over the kinematics shown. See text for explanation of the curves. In both plots, uncertainties shown are statistical only; systematic uncertainties are anticipated to be <4\%. In B, the statistical uncertainties are smaller than the symbols.}
\end{figure}

\vspace{-0.3cm}
\section{Extraction of production time $\tau_p$}
\vspace{-0.2cm}
Space constraints do not permit a full description of the ongoing effort to extract the production time (quasi-free quark lifetime) from a combined fit to $\rm \Delta p_T^2$ and to the hadronic multiplicity ratio, however, the key ideas and results can be given. In a classical geometric model where it is assumed that $\rm \Delta p_T^2$ is proportional to the path length of the quasi-free quark event-by-event, there is a region of kinematics where the production time can be estimated.  The required condition in this picture is that the path length is smaller than the dimensions of the largest nucleus (Pb; $\rm R_{Pb}\approx 6.5~fm$). In that case, the behavior of $\rm \Delta p_T^2$ vs. nuclear radius ~$A^{1/3}$ becomes non-linear, and the degree of non-linearity can be used to estimate the average production length $\tau _L \equiv c \cdot \tau_p$. The model fit can be performed with either 3 or 4 parameters: scale factor, production length, effective total cross section, and optionally, partonic energy loss, which is currently implemented in an approximate way. Production lengths found from this approach currently range from 1.6 to 2.0 fm for the CLAS data, for 4 and 3 parameters, respectively. 

\vspace{-0.3cm}
\section{Exploring a direct measurement of quark energy loss}
\vspace{-0.2cm}
A new idea is under exploration for a direct measurement of quark energy loss from the CLAS data. The basis for the idea is the feature of pQCD energy loss that it is independent of energy as long as the quark path length through the system in question is less than a critical length $L_c$, which is itself energy-dependent. Estimates for the CLAS data suggest that the path lengths through the nuclei are less than the critical length at the energies of these data. If the energy loss is indeed energy-independent, it means that the higher energy part of the pion energy spectrum for the deuterium target should be replicated by a portion of the spectrum from a heavier target at lower pion energies, i.e., there should be a simple {\em{shift}} of part of the deuteron spectrum to lower energies, and the size of this shift is identically the quark energy loss. Thus far, the initial results found for parton total energy loss from a shape analysis of these spectra are consistent with expectations from the pQCD picture. Further studies are required in order to establish the robustness of these studies to a number of minor needed corrections. 

\vspace{-0.3cm}
\section{Conclusions}
\vspace{-0.2cm}
Systematic features of $\rm\pi^+$ $\rm p_T$ broadening have been explored, such as $\rm x_F$ dependence and W dependence. The $\rm x_F$ dependence indicates broadening comes from both current and target fragmentation for $\rm z<0.5$. The 'production time' (virtual quark lifetime) has been extracted from 
$p_T$ broadening data and found to be in the range 1.6-2 fm/c within a geometric model prototype. We have made the first observation of a $\rm \phi_{pq}$ dependence of $\rm p_T$ broadening that increases with target size. This observation suggests that a quantum description of this process is required. Finally, a promising method for directly extracting quark energy loss is being explored, giving results consistent with the observed $\rm pT$ broadening and production time analysis. Experiments of this type can be relevant for the interpretation of jet quenching and proton-nucleus collisions at RHIC and LHC\cite{atlas:2011}\cite{accardi:2011}, and will provide new insight into the fundamental processes involved in QCD hadronization. 

%%%%%%%%%%%%%%%%%%%%%%%%%%%%%%%%%%%%%%%%%%%%%%%%
%% BACKMATTER
%%%%%%%%%%%%%%%%%%%%%%%%%%%%%%%%%%%%%%%%%%%%%%%%

\vspace{-0.3cm}
\begin{theacknowledgments}
\vspace{-0.2cm}
  The authors gratefully acknowledge generous support from Chilean FONDECYT grants 1080564 and 3100064, Chilean CONICYT grant ACT-119, and Chilean BASAL grant FB0821. 
\end{theacknowledgments}

\vspace{-0.3cm}
\bibliographystyle{aipproc}

\end{document}